\documentclass[opre]{informs3nothing}  

\DoubleSpacedXI 

\usepackage{tikz}
\usepackage{gensymb}
\usepackage{multirow}
\usepackage{hyperref}
\usepackage{caption}
\usepackage{listings}
\usepackage{color}
\usepackage{booktabs}
\usepackage{tabularx}
\usepackage{caption}
\usepackage{subcaption}

\usepackage{natbib}
 \bibpunct[, ]{(}{)}{,}{a}{}{,}%
 \def\bibfont{\small}%
\graphicspath{{./figures/}}

\TheoremsNumberedThrough     
\ECRepeatTheorems

\EquationsNumberedThrough    

\TITLE{Mitigating the Backfire Effect Using Pacing and Leading}
\RUNTITLE{Mitigating the Backfire Effect Using Pacing and Leading}
\RUNAUTHOR{Yang, Qureshi, and Zaman}
\ARTICLEAUTHORS{%
\AUTHOR{Qi Yang}
\AFF{%
Massachusetts Institute of Technology,
\EMAIL{yangqi@mit.edu}
}
\AUTHOR{Khizar Qureshi}
\AFF{%
Massachusetts Institute of Technology, Millennium Management,
\EMAIL{kqureshi@mit.edu}
}
\AUTHOR{Tauhid Zaman}
\AFF{%
Yale School of Management,
\EMAIL{tauhid.zaman@yale.edu}
}
}

\SUBJECTCLASS{social networks $|$ social media $|$ political polarization $|$ computational social science} 

\ABSTRACT{
Online social networks create echo-chambers where people are infrequently exposed to opposing opinions. Even if such exposure occurs, the persuasive effect may be minimal or nonexistent.  Recent studies have shown that exposure to opposing opinions causes a backfire effect, where people become more steadfast in their original beliefs. We conducted a longitudinal field experiment on Twitter to test methods 
that mitigate the backfire effect while exposing people to opposing opinions. Our subjects were Twitter users with anti-immigration sentiment. The backfire effect was defined as an increase in the usage frequency of extreme anti-immigration language in the subjects’ posts. We used automated Twitter accounts, or bots, to apply different treatments to the subjects. One bot posted only pro-immigration content, which we refer to as arguing. Another bot initially posted anti-immigration content, then gradually posted more pro-immigration content, which we refer to as pacing and leading. We also applied a contact treatment in conjunction with the messaging based methods, where the bots liked the subjects’ posts. We found that the most effective treatment was a combination of pacing and leading with contact. The least effective treatment was arguing with contact. In fact, arguing with contact consistently showed a backfire effect relative to a control group. These findings have many limitations, but they still have important implications for the study of political polarization, the backfire effect, and persuasion in online social networks.}

\begin{document}

\maketitle

\section{Introduction}
In many situations one may want to persuade an audience to change their  behavior or shift their underlying opinion.  One natural persuasion method is to present an argument to the audience supporting a target opinion or position.  Classical opinion dynamics models  suggest that repeated arguments will cause the audience's opinion to shift toward the target position \citep{degroot1974reaching, demarzo2003persuasion,golub2010naive}.  These models assume that this shift occurs no matter the distance between the opinions of the argument and the audience. More recently, models have been proposed which  assert individuals have bounded confidence, meaning that opinions which differ too much from their own have no persuasive effect \citep{hegselmann2002opinion}.  Empirical research has shown that when opinions differ greatly, making an argument can actually cause the opinions of the audience to shift away from the argument  \citep{lord1979biased, nyhan2010corrections, bail2018exposure}. This \textit{backfire effect} poses a major challenge when trying to persuade or influence individuals.

Today online social networks provide a platform for one to persuade a potentially large audience \citep{perrin2015social}.
However, the structure of these networks present their own obstacles to persuasion.  Because users can choose from whom they receive information, these networks exhibit a great deal of homophily, where neighbors have similar opinions \cite{bakshy2015exposure}.  This creates echo-chambers where users are not frequently exposed to arguments contrary to their own positions and existing opinions are often reinforced.  Within such online settings it has been found that the use of uncivil or extreme language can spread in such online settings  \citep{cheng2017anyone}.  Such language can create animosity among social media users and   prevent constructive discussions. 

The combination of the backfire effect and echo-chambers  present major obstacles to persuasion.  The structure of echo chambers prevent one from seeing contrary opinions, but if one does, the backfire effect limits their persuasion ability.  Given the scale and importance of online social networks, it is important to develop methods to persuade in these environments.  It would be useful to have a method that allows one to present arguments in online social networks in a manner that mitigates the backfire effect and the usage of extreme language.

In this work  we conduct a field experiment to test persuasion methods in an online social network.  Our standard method, which we refer to as \textit{arguing}, simply has one present arguments for the target position without any other interaction with the audience.  Arguing can be viewed as a messaging based persuasion method because it only involves content posted by the arguer.  We test another messaging method we refer to as \textit{pacing and leading} which is based on the idea that persuasion is more effective if there is some sort of bond or connection  between arguer and audience.  This method begins by having the arguer emotionally pace the audience by agreeing with their opinion on the persuasion topic.  This is done to form a bond with the audience.  Then over time, the arguer shifts its own opinion towards the target position which will lead the audience to this position.  In addition to messaging based methods, we also test a persuasion method based on interaction with the audience that we refer to as \textit{contact}.  This method has the arguer like the social media posts of its audience.  This interaction can serve as a form of social contact in an online setting and potentially lead to more effective persuasion when combined with messaging based methods.

Our experiment tests two primary hypotheses.  The first hypothesis is that pacing and leading will mitigate the backfire effect more than standard arguing through the effect of in-group membership, which means that the arguer and audience belong to a common social group.  Theories of inter-group conflict suggest that persuasion is more effective when the arguer and audience are in-group \cite{tajfel1979integrative}.      In \citep{munger2017tweetment} race was used as an in-group feature to persuade users in the online social network Twitter to not use extreme language.  It was found  that in-group persuasion (arguer and audience have the same race) was more effective than out-group persuasion (arguer and audience have different races).  This study demonstrated that race was an effective in-group feature for persuasion.  We expect a similar finding when in-group membership is based the opinion towards the persuasion topic.

Our second hypothesis is that contact between the arguer and audience will mitigate the backfire effect. By having contact with the audience,  the arguer can form a rapport with the audience and shift them to a more positive affective state.  Persuasion strength may be enhanced by these psychological effects.  Researchers have found that affective states impact the efficacy of persuasion \citep{rind1997effects,rind2001effect}. The social influence literature is rife with evidence that social rapport and a positive relationship  enhance persuasion and influence \citep{cialdini1998social}. Moreover, it has been found that a person's persuasive ability is strengthened if the audience likes this person \citep{burger2001effect}.

\section{Experiment Design}
\begin{figure*}
	\centering
	\includegraphics[scale = 0.75]{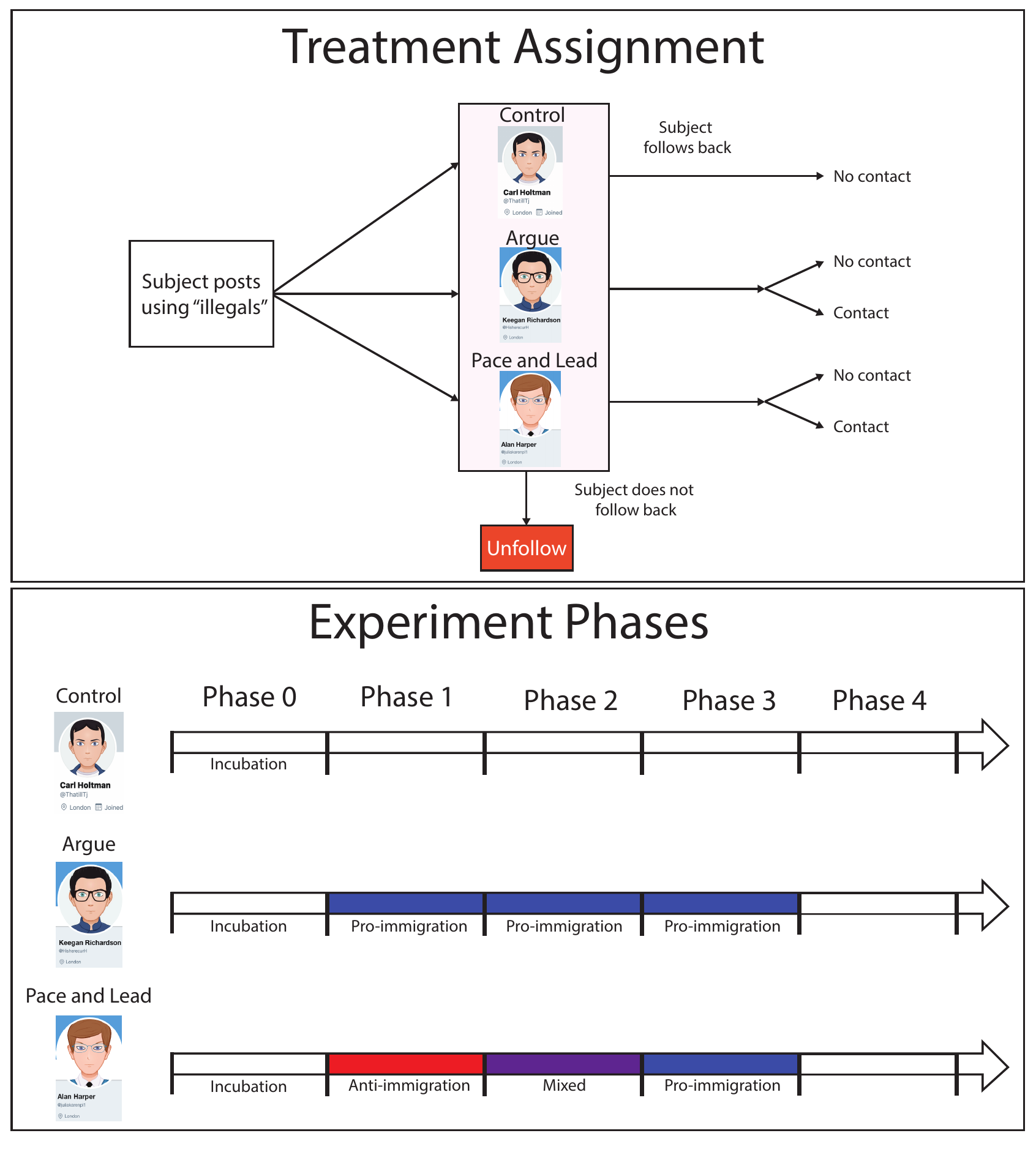}
	\caption{(top) Diagram illustrating the subject acquisition procedure for the experiment.  There is one bot for each tweet based treatment: control, argue, and pace and lead. The pool of subjects consists of all users who posted tweets with the word 	``illegals''.  These users are randomly assigned to the bots.  To acquire experiment subjects the bots like a tweet of their assigned users and follow them.  Users who follow-back become experiment subjects and users who do not follow back are unfollowed by the bots.  We then randomly assign the subjects of each bot to either the no contact or contact group.  Subjects in the contact group will have the bot like at most one of their tweets per day if they post a tweet.   The control bot does not apply the contact treatment.		
	(bottom) Timeline of experiment phases. Phase zero is the incubation period where the bots  post neutral content in order to appear human.  The control bot posts no content in phases one to four.  The argue bot posts pro-immigration content in phases one to three and posts nothing in phase four. The pacing and leading bot posts anti-immigration content in phase one, mixed opinion content in phase two, pro-immigration content in phase three, and nothing in phase four.    }
	\label{fig:design}
\end{figure*}

The persuasion topic used in our study is  immigration.  Events such as the European refugee crisis have made immigration a charged political issue and it is an active topic of discussion on social networks.  Several studies have measured population level sentiment on this topic in  Twitter \cite{ozturk2018sentiment,backfried2016sentiment,coletto2016sentiment}.  It was found in \cite{ozturk2018sentiment} that English posts about the refugee crisis were more likely to have a negative opinion on the topic.    A similar result was found for Twitter users in the United Kingdom \cite{coletto2016sentiment}.  Given the level of interest in the topic and its geo-political importance, immigration is an ideal topic to test persuasion methods.  In our experiment we try to persuade individuals to have a more positive opinion of immigration.

We employ automated Twitter accounts, which we refer to as bots, to test different persuasion methods. Our experiment subjects are Twitter users who actively discuss immigration issues and have anti-immigration sentiment. Each bot implements a different persuasion method. One bot is a control which posts no content and does not interact with the subjects. One bot applies the arguing method by posting content which is  pro-immigration. The third bot applies pacing and leading by posting content that is initially anti-immigration and then gradually become more pro-immigration. To test the contact treatment, we randomly selected half of the subjects from each bot and have the bots like the posts of these subjects. To assess the effectiveness of the different persuasion methods, we analyze the sentiment of content posted by these subjects over the course of the experiment. We now present details of our experiment design, which is illustrated in Figure \ref{fig:design}.

The subjects for our experiment were Twitter users who have an anti-immigration sentiment. To find potential subjects we began by constructing a list of phrases that conveyed strong anti-immigration sentiment, such as  \#CloseThePorts, \#BanMuslim, and \#RefugeesNotWelcome. The complete list of phrases is provided in \emph{SI Appendix}.  We used the Twitter Search API \cite{rest-search} to find posts, known as tweets in Twitter, that contained at least one of these keywords.  We then collected the screen names of the users who posted these tweets.   

Our search procedure has the potential to find users who do not have anti-immigration sentiment.  For instance, to convey support for immigrants, a user could post a tweet critical of an anti-immigration phrase. To make sure that there were not many users who fall in this category, we manually investigated 100 random users collected by our search procedure.  We found that none of the users was pro-immigration, giving us confidence that the overwhelming majority of our potential subjects were anti-immigration.

We further narrowed our subject pool by requiring each user to satisfy the following criteria. First, their tweet must be in English and must not contain only punctuation or emojis. Second, the user should not be an automated bot account. The text conditions on the tweet were checked using simple pattern matching. Bot accounts were identified using the machine learning based Botometer algorithm \cite{davis2016botornot}. Users who Botometer identified as being the most bot-like were manually reviewed and eliminated if they are indeed bots.

We created Twitter accounts for the control, argue, and pace and lead treatments.  One of the goals of our experiment was to test persuasion strategies in a realistic setting.  Therefore, we wanted the bots to resemble human Twitter users, in contrast to the study in \cite{bail2018exposure} where the subjects were told in advance the Twitter account they were following was a bot.  To accomplish this, we had the bots be active on Twitter for a two month incubation period before we started the experiment.  Each of the bots location was set to London, and they followed a number of popular British Twitter accounts.    The bots were designed to look like white males with traditional European names.  We used cartoon avatars for the profile pictures, similar to what was done in \cite{munger2017tweetment}.
We show the profile images for the bots and list their treatment type in Figure \ref{fig:design}.     During the incubation period the bots posted tweets about generic, non-immigration topics and shared tweets about trending topics on Twitter, an act known as retweeting.  They also tweeted  articles or videos talking about immigration, but not yet taking a stance on the issue.  This was done to show that the bots had some interest in immigration before the experiment began. During the incubation period the bots tweeted once or twice a day. 
 We provide examples of the incubation period tweets and retweets in \emph{SI Appendix}.

One month into the incubation period, we began obtaining subjects for the experiment.  To participate in the experiment, the potential subjects needed to follow the bots.  This way they would be able to see the tweets posted by the bots in their Twitter timelines.   We randomly assigned each of the users in the subject pool to the bots.  The bots then liked a recent tweet of their assigned users and followed them.  The liking of the tweet and following were done to increase the follow-back rate of the potential subjects.  To avoid bias before the experiment, all tweets the bots liked were manually verified to not be immigration related.  After liking and following their assigned subjects' tweets, the bots were able to achieve an average follow back rate of 19.3\%. In total we were able to obtain 1,336 subjects who followed the bots. To make the bots appear more human, we tried to keep their ratio of followers to following greater than one. To do this, the bots would wait one to seven days before unfollowing a user who did not follow-back.  The actual wait time depended on the user activity level, with a longer wait time given for less active users. Details are provided in \emph{SI Appendix}. 

%



The experiment had four different phases. We denote the incubation period as phase zero.  Phases one, two, and three are the main active phases of the experiment. 
The control bot does nothing for these phases. The argue bot would post a pro-immigration tweet once a day in these phases. The pace and lead bot also posted tweets once a day in these phases, but the tweet opinion varied. In phase one the tweets were anti-immigration.  In phase two the tweets expressed uncertainty about immigration or potential validity of pro-immigration arguments.  In phase three the tweets were pro-immigration, similar to the argue bot.  We constructed the tweets based on what we deemed a proper representation of the opinion for each phase.
We show example tweets for the argue and pace and lead bots in the different phases in \emph{SI Appendix}. In phase four of the experiment the bots tweeted nothing.  We used this phase to measure any persistent effect of the treatments.  Each phase lasted approximately one month, except for the incubation phase which lasted two months. The incubation phase began on September 27th, 2018 and the fourth phase was completed on March 1st, 2019.  The experiment timeline is shown in Figure~\ref{fig:design} and precise dates for the phases are provided in Table~\ref{table:phase_coef}.

 In addition to the tweeting based treatments, we also tested the contact treatment on the subjects.  We randomly assigned 50\% of the subjects of the argue and pace and lead bots to this treatment group.  During phases one, two, and three, the bots liked the tweets of the subjects assigned this treatment.  When the bot liked a subject's tweet, the subject is notified.  Liking tweets would make the bot more visible to the subject and potentially give the subject a greater trust or affinity for the bot.  The control bot did not apply the contact treatment to any of its subjects.

 All subjects  voluntarily chose to follow the bots.  Those who did not follow the bots potentially represent a different type of subject.  Therefore, our conclusions are limited to Twitter users willing to follow the bots and do not necessarily generalize to all Twitter users.  However, since a follow-back is required for a Twitter account to implement a tweet based treatment, this is not a strong limitation of our conclusions.   This research was approved by COUHES, the Institutional Review Board (IRB) for the Massachusetts Institute of Technology (MIT).


\section{Results}
\begin{figure*}
	\centering
	\includegraphics[scale = 0.12]{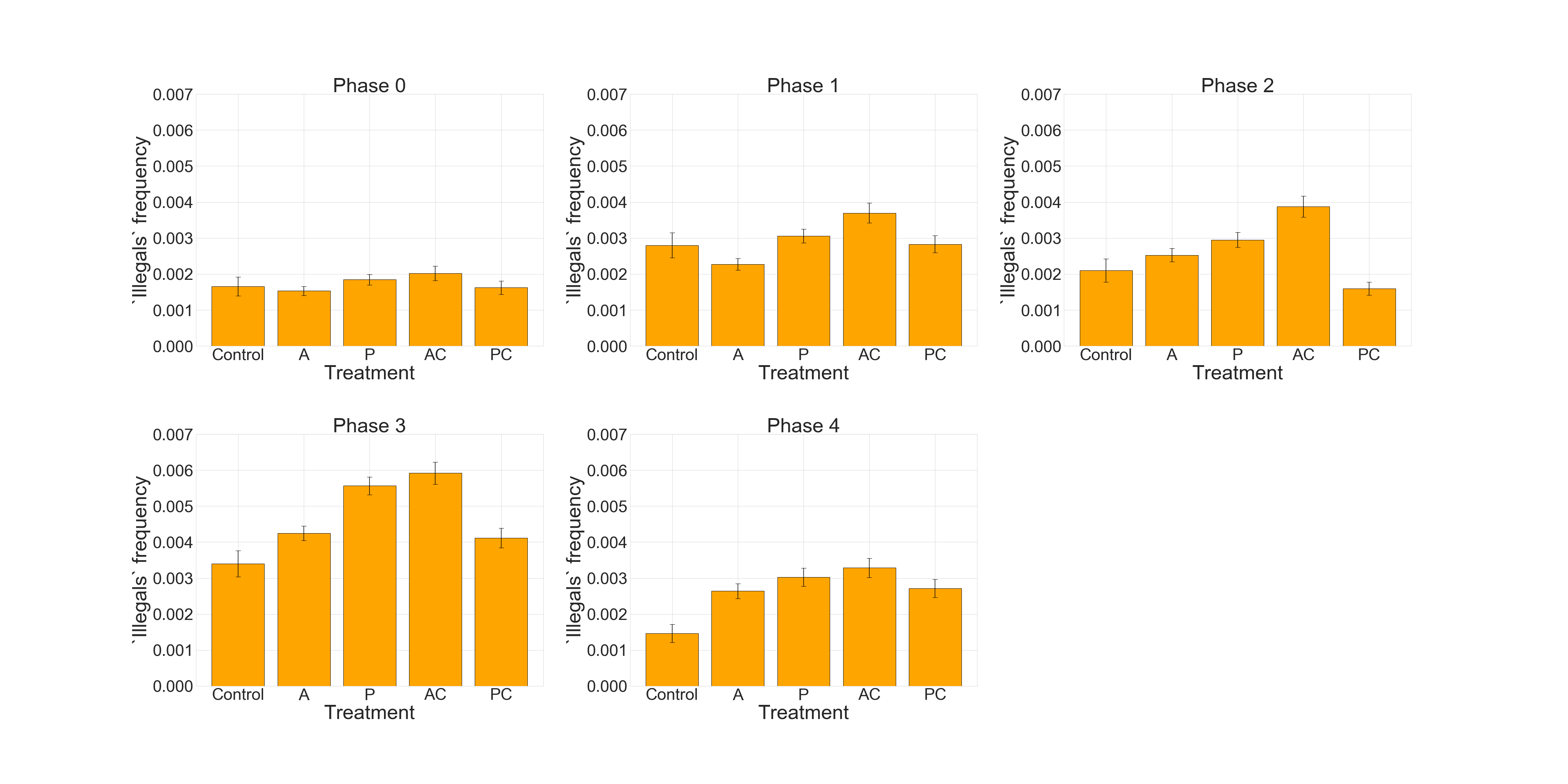}
	\caption{Plot of the frequency and standard error of usage of the word ``illegals'' in tweets for each phase and treatment group.  The frequency is calculated as the total number of tweets containing illegals divided by the total number of tweets for all subjects in the treatment group during the corresponding phase.   The treatments are labeled as follows: A is argue without contact, AC is argue with contact, P is pace and lead without contact, and PC is pace and lead with contact.  }
	\label{fig:illegals_freq}
\end{figure*}

We used the frequency of extreme anti-immigration language in the subjects' tweets to measure any persuasion effect the bots had.  In particular,  we counted how many of the subjects' tweets contained the word ``illegals'' in each phase.  The term illegals is a pejorative term used by people with anti-immigration sentiment.   For instance, there are tweets such as \textit{I want a refund on all the tax money spent on illegals!!!} which show strong anti-immigration sentiment. The usage frequency of such extreme language can be used to gauge sentiment, as was done in \cite{munger2017tweetment}.  We chose the word illegals because it is consistently used by anti-immigration Twitter users, unlike hashtags that gain temporary popularity.    We plot the illegals usage frequency in each phase and treatment group in Figure \ref{fig:illegals_freq}.  This frequency is defined as the number of tweets containing illegals divided by the total number of tweets for all subjects in each phase and treatment group.  We note that the overall frequency is very low, but shows aggregate differences between phases.  For instance, phase three has a higher frequency than the other phases for all treatments.  This suggests that there are exogenous factors affecting the behavior of the subjects.  Another interesting observation is in phase two, where we see that the pace and lead with contact treatment has a much lower frequency than the other treatments, while argue with contact has the highest frequency.  Recall that in phase two pacing and leading has tweets that are slightly pro-immigration.  We next perform a more quantitative statistical analysis  to assess the different treatments.

We treat each tweet as a binary outcome that equals one if the tweet contains the word illegals.  The probability of such an outcome is modeled using logistic regression.  For a tweet $i$ the  probability is
\begin{align}
\log\frac{p_i}{1-p_i} = &  \sum_{t=0}^4\beta_{t}x_{t,i} + 
\sum_{t=0}^4\beta_{a,t}x_{a,i}+\nonumber\\
& \sum_{t=0}^4\beta_{p,t}x_{p,i}+
\sum_{t=0}^4\beta_{ac,t}x_{ac,i}+\nonumber\\
& \sum_{t=0}^4\beta_{pc,t}x_{pc,i} + \epsilon_i.\label{eq:control}
\end{align}
The coefficients $\beta_t$ for $t=0,1,..4$ model exogenous factors that may impact the probability during each phase.  For instance, news stories related to immigration may increase the probability.  We use separate treatment coefficients for each phase because the pace and lead treatment varies by phase. Recall that this treatment shifts the opinion of its tweets from anti- to pro-immigration over phases one to three.  The treatment coefficients are indexed by subscripts indicating the treatment and phase.  We use the subscript $t$ for the phase, $a$ for argue, and $p$ for pace and lead.  The subscript  $c$ indicates the contact treatment where the bots like the subjects' tweets.   The $x$ variables are binary indicators for the treatment group of the subject posting the tweet and in which phase the tweet occurred.
User heterogeneity and other unobserved factors are modeled using a zero mean normally distributed random effect $\epsilon_i$.

\begin{table}	
	\centering
	\begin{tabular}{ |c|c|c| } 
		\hline
		Phase & Dates&  Coefficient (p-value)   \\ 
		&            &                \\
		\hline
		0 & 2018-09-27 to 2018-10-27 & -6.40 (0.001)\\ \hline
		1 &2018-10-27 to 2018-11-27&  -5.87 (0.001) \\ \hline
		2 &2018-11-27 to 2018-12-25&  -6.16 (0.001) \\ \hline
		3 & 2018-12-25 to 2010-01-29 &  -5.67 (0.001)\\ \hline     
		4 & 2019-01-29 to 2019-03-01 &  -6.52 (0.001)\\ \hline
	\end{tabular}
	\caption{Dates of each phase and regression coefficients with p-values for phase effects.}
	\label{table:phase_coef}
\end{table}

 We show the regression coefficients for the phase term along with the dates for each phase in Table \ref{table:phase_coef}.    We see that the phase effect is not constant and peaks in phase three.  We note that a larger coefficient value implies an increased usage frequency of illegals.  Upon further investigation we found that during this time period the U.S. government was shutdown because Congress would not provide President Donald Trump funding for a border security wall.
We suspect this caused an overall increase in the usage frequency of the word illegals by anti-immigration users on Twitter.

By regressing out the phase effect we can isolate the different treatments. We plot the resulting treatment coefficients separated by tweet group (argue or pace and lead) and contact group in  Figure \ref{fig:coef}.  This grouping makes differences in each individual treatment over the phases more visible.  We also indicate on the plots which differences  are statistically significant at a 1\% level.  The significance levels for the differences are calculated by repeating the regressions using different treatments as the reference group.

We first look at the effect of the contact treatment.  In the top left plot of Figure \ref{fig:coef} we see that the argue with contact coefficient is greater  than argue without contact,  and the difference does not vary much over the phases.  The difference is significant for phases one, two, and three.  In phases zero and four, where the bots do not tweet about immigration, there is no significant difference.    The contact treatment may be making the bots' pro-immigration tweets more visible to the subject, resulting in a backfire effect where the subject uses the word illegals more frequently.     

For pacing and leading in the top right plot of Figure \ref{fig:coef}, we see that the non-contact coefficient is greater than contact. In phases two and three the difference is significant.  Contact appears to enhance the effectiveness of pro-immigration tweets in the later stages of the pacing and leading treatment.   This is in contrast to arguing, where contact degrades the effectiveness of pro-immigration tweets.       

We next look more closely at arguing versus pacing and leading when the contact treatment is fixed.  In the bottom left plot of Figure \ref{fig:coef} we see that without contact, the tweet treatment coefficients have a small difference which does vary appreciably across phases.  Argue has a smaller coefficient, but the difference is statistically significant only for phases one and three.    

For the contact group in the bottom right plot of Figure \ref{fig:coef}, the difference changes sign.  Argue has the larger coefficient and the difference varies across the phases.  Phase two shows a large significant difference.     The difference is smaller in phase three, but still significant.  The moderately pro-immigration tweets of the phase two pace and lead treatment seem to be more effective than the argue tweets when the bot has contact with the subject.  The same can be said of fully pro-immigration tweets in phase three, but the advantage of pacing and leading over arguing is less than in phase two.

\begin{figure*}
	\centering
	\includegraphics[scale = .12]{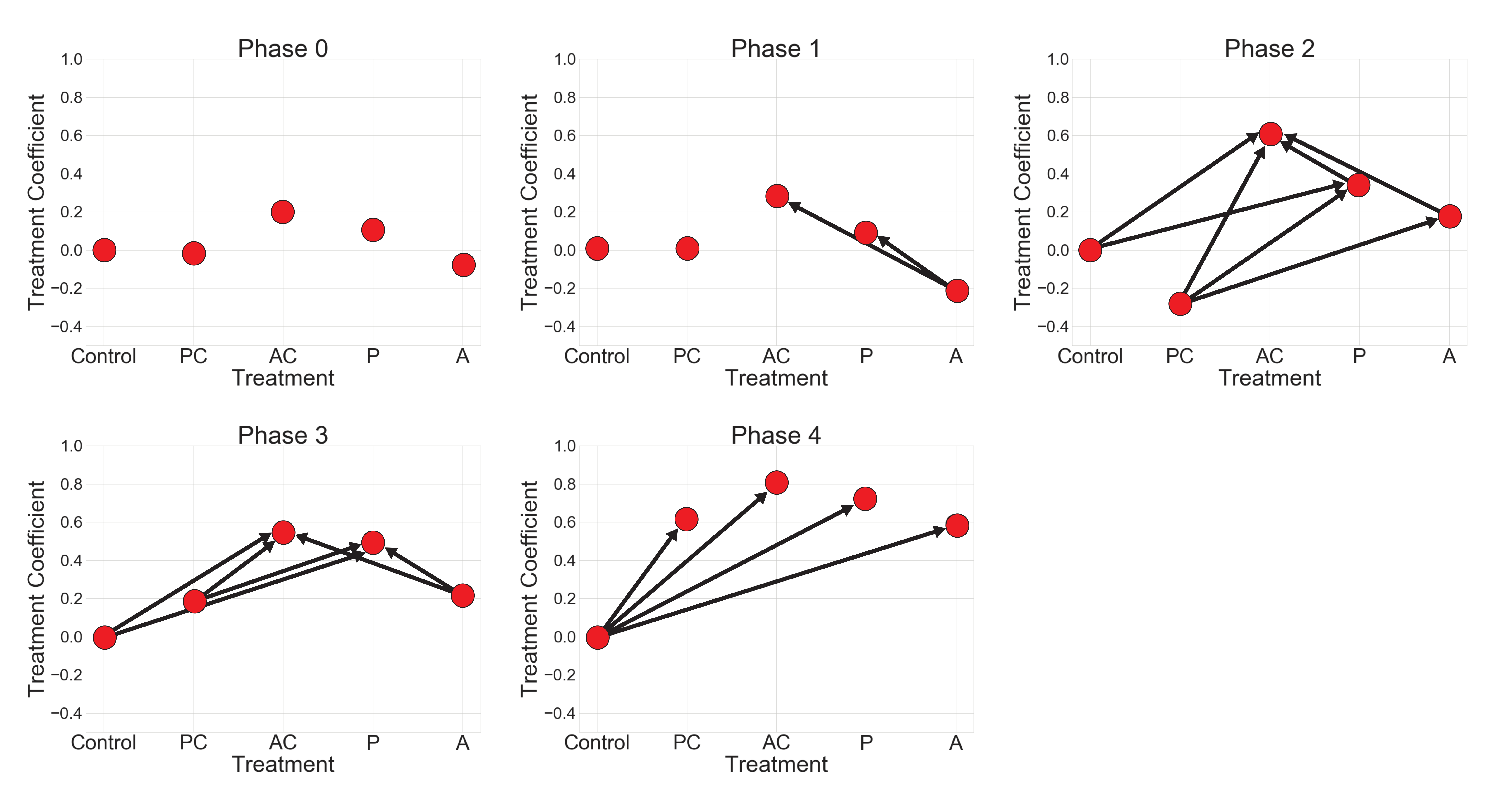}
	\caption{Plot of the treatment regression coefficients grouped by phase.  Edges are drawn indicating which treatments were more effective at reducing the extreme language usage rate in each phase of the experiment.  An edge is drawn between two coefficients if their difference  is statistically significant at a 1\% level.  The treatments are labeled as follows: A is argue without contact, AC is argue with contact, P is pace and lead without contact, and PC is pace and lead with contact.    }
	\label{fig:dags}
\end{figure*}

The results in Figure \ref{fig:coef} suggest that phase two pacing and leading with contact is the most effective treatment.  To make this more precise,
we plot the treatment coefficients grouped by phase in Figure \ref{fig:dags}.  The control group has a coefficient of zero, which we indicate in the figure.
We also place directed edges between treatments that have a statistically significant difference at a 1\% level.  This creates a directed acyclic graph (DAG)
for the treatments.    These treatment DAGs more clearly show which treatments are most effective in each phase.  

We see that in phase zero there are no edges, indicating that each treatment group begins at roughly the same state.
In phase one,  argue without contact is better than pace and lead without contact.   Also we see that contact makes arguing
less effective, as we saw in Figure \ref{fig:coef}.  

In phase two we start seeing many significant differences $(p<0.01)$ between treatments.
Pacing and leading with contact is the most effective treatment, with edges pointing  from it to the three other treatments.
Arguing with contact is the worst treatment.  Arguing with contact and pacing and leading without contact have an edge
from control, which indicates that they cause a backfire effect relative to doing nothing. 

In phase three the best treatments are argue without contact and pace and lead with contact.  These two treatments have no
significant difference between each other.  Argue with contact and pace and lead without contact  both show a backfire effect relative to control.
The backfire effect of argue with contact  may be due to the contact making the pro-immigration tweets more visible.  For pace and lead
it is not clear why a backfire effect occurs in the absence of contact.  In phases two and three  the pace and lead treatment posts tweets that support immigration to different degrees.  Without the contact treatment,
going from anti-immigration to  pro-immigration causes a backfire effect in the subjects.  It may be that the combination of pacing (anti-immigration tweets) and contact creates a sense of trust with the subjects that prevents the backfire effect when the tweets become pro-immigration.  When only pacing is done, there may not be enough trust, and the switch in immigration sentiment of the treatment causes a backfire.

In phase four we see that all four treatments
show no significant differences between each other.  However, they all show a backfire effect.   It is not clear why this backfire occurs.  
Nevertheless, we do see that the treatments have no persistent differences between each other once they are terminated.

\begin{figure*}
	\centering
	\includegraphics[scale = 0.4]{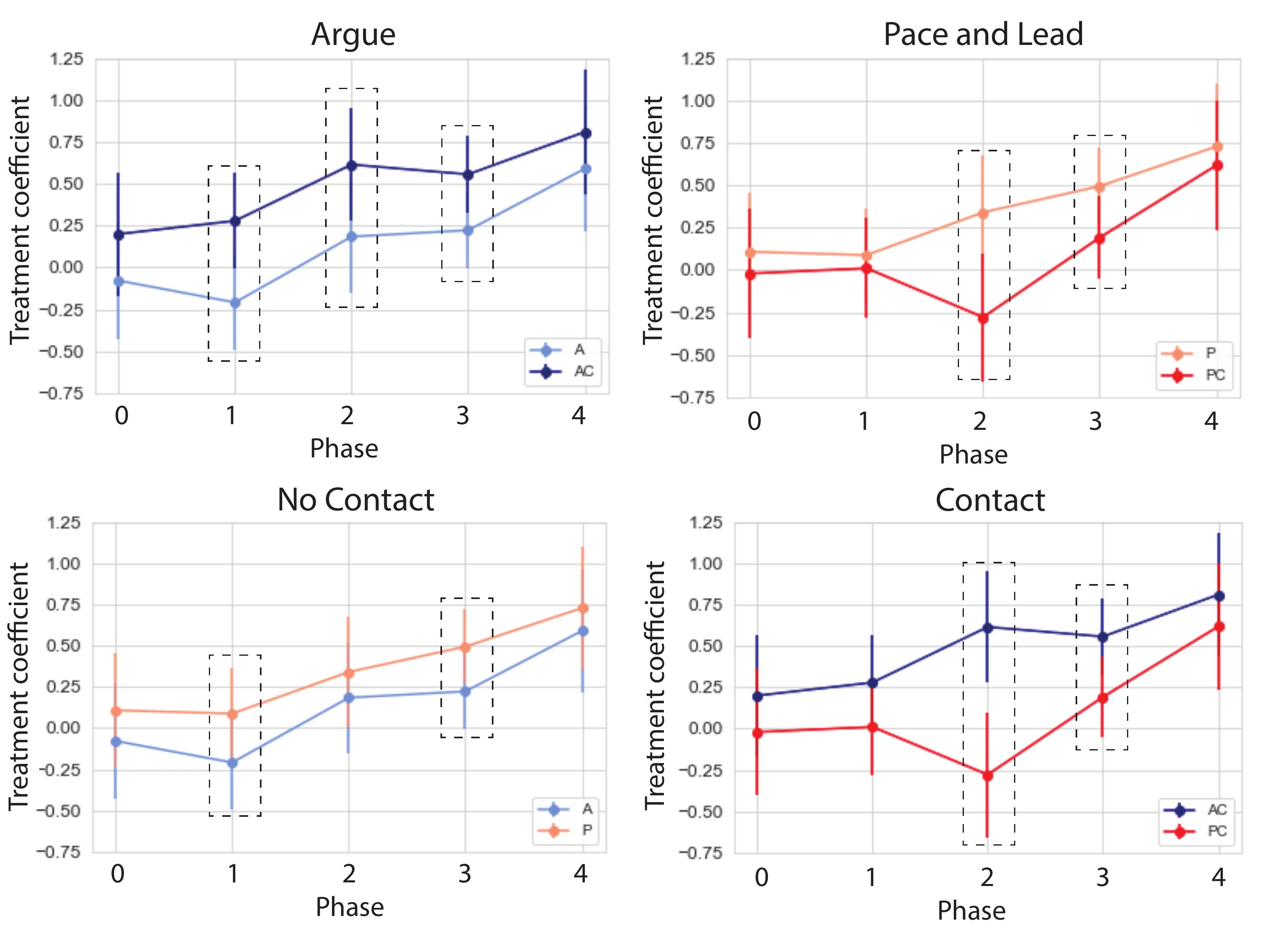}
	\caption{Plots of the regression coefficients (with standard errors) for the tweet and contact treatments versus phase.  The title of each plot indicates the treatment component that is held fixed.  In the legend A is argue without contact, AC is argue with contact, P is pace and lead without contact, and PC is pace and lead with contact.   The dashed boxes indicate which coefficients have a difference that is statistically significant at a 1\% level. }
	\label{fig:coef}
\end{figure*}

\section{Discussion and Conclusion}

Our original hypotheses concerned which treatments would mitigate the backfire effect, which we define as an increase
in the use of extreme anti-immigration language in tweets.  We hypothesized that pacing and leading would be more effective than arguing and that contact would be more effective than no contact.  Our results present a more complex situation. With contact, pacing and leading was more effective than arguing in phase two and three.  For arguing, contact made the treatment less effective, while for pacing and leading, contact improved the the treatment.  Therefore, our findings indicate that our  binary hypotheses do not encompass the nature of the treatments.  Rather, we see a novel interaction effect, where combining pacing and leading with contact is the most effective treatment in phase two.

The challenges of echo-chambers and the backfire effect make persuasion non-trivial in online social networks.  Our findings suggest strategies
one can use to overcome these challenges.  Echo-chambers can be overcome if one can penetrate the local network of a social media user.
We were able to accomplish this using bots which followed and liked posts of the user.  This follow and like method proved to have a high rate of getting users to follow the bots.  Penetrating a user's network allows one to present arguments to the user.  However, the backfire effect would result in these arguments making the user more steadfast in their original belief.  We found that to overcome the backfire effect, the bot should continuously like the posts of the user, and present arguments that are more nuanced and moderate in their language.  This softer approach proved more effective than standard arguing.
Our study presents new applied techniques to run influence campaigns in online settings  across an extended time period. These techniques represent an important advance for the field of computational social science \cite{lazer2009computational}.

Our findings also present many interesting questions that merit further investigation. One question concerns the phases for pacing and leading. We found that the moderate posts were most effective. It is not clear if this treatment would work in isolation or if the phase one pacing treatment is necessary.  We hypothesize that this period allows greater trust to be built between subject and bot, but our experiment does not confirm this.  Another question is whether the phase three pace and lead treatment where the posts strongly advocate the target position is necessary.  It may be that the moderate posts are sufficient to mitigate the backfire effect and potentially persuade the subject.  


Care should be taken when trying to generalize our findings to more general settings.  Our study focused on the topic of immigration which is an important political and policy issue.  Discussion on this topic has split along traditional conservative liberal fault lines.  We expect our findings to extend to similar political issues, but further study is needed.   However, our subjects were Twitter users with anti-immigration sentiment who were willing to follow our bots.  This represents a limited population in a very specific social setting.  More work is needed to determine whether our findings replicate in different populations or within varied social settings.

\ACKNOWLEDGMENT{The authors would like to thank Jeremy Yang, Dean Eckles, and Sinan Aral for helpful discussions. }

\bibliography{pnas-sample}
\bibliographystyle{plainnat}

\newpage
\APPENDIX{}
\section{Keyword for Subject Acquisition}\label{sec:keywords}
We show in table \ref{table:keywords} the keywords used to find experiment subjects.  We used the Twitter Search API to find tweets containing the keywords and the users posting the tweets become potential subjects.

\begin{table}[h!]
\centering
\begin{tabular}{@{}ll@{}}
\toprule

1       RefugeesNotWelcome & 12               StopIslam \\
2               Rapefugees & 13            ISLAMIZATION \\
3               BanMuslims & 14         UnderwearBomber \\
4            WhiteGenocide & 15             NoRefugees  \\
5            StopRefugees & 16    StopIllegalMigration \\
6           CloseThePorts & 17         AntiImmigration \\
7     ImmigrationInvasion & 18           Reimmigration \\
8             MigrantCrime & 19              NoRefugees \\
9                FreeTommy & 20                 NoIslam \\
10 QAnon & 21        ProtectOurBorder \\
11                   MAGA \\\bottomrule
\end{tabular}
\caption{Hashtags used to identify target users}
\label{table:keywords}
\end{table}

\section{Example Bot Tweets}\label{sec:bottweets}
The experiment has four phases numbered zero to three.  Phase zero is the incubation period where the bots post content which does not take a stance on immigration.  The argue bot posts pro-immigration tweets in phases one, two, and three.  The pace and lead bot posts anti-immigration tweets in phase one. In phase two its tweets express uncertainty about immigration or potential validity of pro-immigration arguments. In phase three the tweets are pro-immigration, similar to the argue bot. We constructed the tweets based on what we deemed a proper representation of the opinion for each phase.

%

Tables \ref{table:ph0},\ref{table:ph1}, and \ref{table:ph3} shows randomly selected examples of the tweets posted by each bot in each phase of the experiment.
\begin{table}[]
\centering
\begin{tabular}{|l|l|l|}
\hline
Phase   & \multicolumn{2}{l|}{Argue Bot and Pace and Lead Bot}                                                                                                                                                 \\ \hline
Phase 0 & \multicolumn{2}{l|}{What an incredible experience \#RyderCup18}                                                                                                                                      \\ \hline
Phase 0 & \multicolumn{2}{l|}{\begin{tabular}[c]{@{}l@{}}Newcastle become the first team in \#PL history to score \\ twice against Man Utd at Old Trafford in the opening \\ 10 minutes \#MUNNEW\end{tabular}} \\ \hline
Phase 0 & \multicolumn{2}{l|}{\begin{tabular}[c]{@{}l@{}}GOAAALLLL!! Shaqiri again playing a big part in \\ the goal. Salah with a smashing finish to make it two!\end{tabular}}                               \\ \hline
Phase 0 & \multicolumn{2}{l|}{Looking forward to Saturday already! \#MondayMotivation}                                                                                                                         \\ \hline
\end{tabular}
\caption{Tweets posted by the bots in phase zero of the experiment.}
	\label{table:ph0}
\end{table}

\begin{table}[]
\begin{tabular}{|p{1.3cm}|p{7cm}|p{8cm}|}
\hline
Phase   & Argue Bot                                                                                                                                                                                        & Pace and Lead Bot                                                                                                                                                                                                                                                 \\ \hline
Phase 1 & Former Calais Jungle child refugee who was unlawfully refused safe passage to join his aunt in Britain still in France two years from the closure of the camp. Can we reunite him with his aunt? & Immigrants strike again. Muslim Uber driver Khaled Elsayedsa Ali charged in California with kidnapping four passengers. This needs to be stopped.                                                                                                                 \\ \hline
Phase 1 & Unbelievable. A revised estimate of 56,800 migrants have died/gone missing over the past four years.                                                                                             & Muslims attempt to derail high-speed train in Germany using steel wire. Threats in Arabic were found thereafter.                                                                                                                                                  \\ \hline
Phase 1 & A win for refugees! Former refugee elected to US congresswoman.                                                                                                                                  & Unacceptable. After mass Muslim migration into Germany, sex attacks are up 70\% in Freiburg alone.                                                                                                                                                                \\ \hline
Phase 2 & Chancellor Angela Merkel defends UN migration pact. A step in the right direction.                                                                                                               & UK Government to sign UN migration pact. Interesting that Angela Merkel defends it, and rejects "nationalism in its purest form". I believe in her.                                                                                                               \\ \hline
Phase 2 & It's human rights day, and refugees across Europe face widespread human rights violations. Europe needs to do more to uphold natural human rights.                                               & "Muslim imam performed call to worship during a Church of England cathedral's Armistice without permission." Crossed the line. However, it would probably be overlooked if it were the other way around, am i right?".                                            \\ \hline
Phase 2 & Now that's efficient and socially productive! Germany sets out new law to find skilled immigrants.                                                                                               & The UN migration pact, which would criminalize criticism of mass migration and redefine a refugee, will be signed by world leaders next week." Though not through public consent, the \#ImmigrationMatters initiative did deliver guiding messages to the public. \\ \hline
\end{tabular}
\caption{Tweets posted by the bots in phases one and two of the experiment.}
	\label{table:ph1}
\end{table}

\begin{table}[]
\centering
\begin{tabular}{|l|l|l|}
\hline
Phase   & \multicolumn{2}{l|}{Argue Bot and Pace and Lead Bot}                                                                                                                                                                                                                                           \\ \hline
Phase 3 & \multicolumn{2}{l|}{\begin{tabular}[c]{@{}l@{}}Pathetic. At the height of the Syrian refugee crisis in 2015, Syria‘s neighbors \\ took in 10,000 refugees per DAY. Yet the UK Home Secretary just called the \\ arrival of 75 asylum seekers by boat in 3 days a major incident.\end{tabular}} \\ \hline
Phase 3 & \multicolumn{2}{l|}{\begin{tabular}[c]{@{}l@{}}Appalling? In 2018 at least 2,242 people have died in the Mediterranean Sea \\ trying to reach Europe.\end{tabular}}                                                                                                                            \\ \hline
Phase 3 & \multicolumn{2}{l|}{\begin{tabular}[c]{@{}l@{}}The sole survivor said he was left alone in the water for at least 1 day before \\ a fishing boat found and rescued him.\end{tabular}}                                                                                                          \\ \hline
\end{tabular}
\caption{Tweets posted by the bots in phase three of the experiment. Both bots tweeted pro-immigration tweets.}
	\label{table:ph3}
\end{table}

\section{Bot Operation}
The bots were active for two months before we started the experiment. We did this to make the bots seem real so that the targets would not be suspicious that they were being followed by a fake account. Their activity included having them first tweet some manually created messages. We also looked at trending topics and retweeted some of those posts, such as UEFA Europa League (we provide more example tweets in table \ref{table:ph0}). Each of the bots' locations were set to London, and they followed a number of common English Twitter accounts to give them the indication of living there.

The bots  started to follow the users we identified as anti-immigration people to gain followers. We made sure that no two bots were following the same user as this could arouse suspicion. To boost the follow back rate, the bots liked the users' tweets. To avoid bias before the experiment, all tweets the bots liked were not immigration related. The bots also unfollowed the users after some given time to prevent our following count from being inflated, and to keep a better ratio of followers to following which is desirable for appearing human and gaining followers. The “unfollow time” depends on user tweet frequency  and was calculated in the following way. Let $W_u$ denote how long the bot waits between following and unfollowing user $u$.   The wait time should reflect how often a user checks Twitter and it should be shorter for more active users because we want to give the user time to log in and see that the bot had interacted with and followed them and then make the choice on whether or not to follow it.  Also, we want the bot to wait at least one day before unfollowing and at most seven days to ensure that it would not wait too long or unfollow too soon.  Let $\mu_u$ and $\sigma_u$ be the mean and standard deviation of the inter-tweet time for user $u$.  Small values for $\mu_u$ indicate that $u$ is a active Twitter user and checks the app often.    Then $W_u$ is given by

\begin{equation}
W_u = \min(7\text{~days}, \max(1\text{~day}, \mu_u + 4 \sigma_u))
\end{equation}


The bot would unfollow the user if the user was not following the bot when the wait time had elapsed.  Users who followed the bot were not unfollowed and became subjects for the experiment.   For the remaining phases of the experiment all tweets from the bot would appear on their Twitter timeline. 

Table \ref{table:followback} shows the number of users each bot attempted to follow and the number of users who followed back and were available throughout the experiment. Users may not be available due to three reasons:
(i) privacy settings, (ii) account deletion by user, (iii) account suspension by Twitter.
\begin{table}[]
\centering
\begin{tabular}{@{}lllll@{}}
\toprule
Bot               & Treatment             & Followed & Followed Back & Available \\
\midrule
Alan Harper       & White, Pacing/Leading & 3045     & 636           & 578       \\

Keegan Richardson & White, Arguing        & 3051     & 717           & 651       \\

Carl Holtman      & White, Control        & 817     & 125           & 107       \\
\bottomrule
\end{tabular}
\caption{Number of users who were followed by, followed back, and remained available for all phases of the experiment for each bot.}
\label{table:followback}
\end{table}
\section{Covariate Balance Check}
Table \ref{table:feature} shows the followers and friend count of the study population. We performed a pair-wise t-test for all groups and we found that there is no statistically significant difference between any group means $(p<0.05)$.

\begin{table}[h]
\centering
\begin{tabular}{lrrrr}
\toprule
{} &  \multicolumn{2}{l}{Followers Count} & \multicolumn{2}{l}{Friends Count}  \\
{} &             mean &    std &            mean &   std  \\
BOT                    &                  &        &                 &          \\
\midrule
A &  6854 &  18510 &  6793 &  12962  \\
AC & 5184 &  9738 &  5622 &  9891  \\
P &  6057 &  11368 &  6155 &  9992 \\
PC & 4754 &  20319 &  4621 &  13950 \\
Control & 6367 &  7860 &  6474 &  7645  \\
\bottomrule
\end{tabular}
\caption{Descriptive characteristics of study population for each bot.  The bot are labeled as follows: Control is the control bot, A is argue without contact, AC is argue with contact, P is pace and lead without contact, and PC is pace and lead with contact.}
\label{table:feature}
\end{table}

\section{Experiment Data}\label{sec:data}
We show in table \ref{table:tweetcount} the number of tweets and number of tweets with the word ``illegals'' in each phase and treatment group.

\begin{table}	
	\centering
		\begin{tabular}{ |c|c|c|c| } 
			\hline
			Phase & Treatment & Number of tweets & Number of tweets   \\ 
			&            &                &containing ``illegals''\\
			\hline
			0 & Control & 24,156 & 40\\ \hline
			0 & Argue& 97,277 & 149\\ \hline
			0 & Pace& 86,236 & 159\\ \hline
			0 & Argue contact & 50,474 & 102\\ \hline     
			0 & Pace contact& 47,467 & 77\\ \hline
			\hline
			
			1 & Control & 23,212 & 65\\ \hline
			1 & Argue& 85295 &  194\\ \hline
			1 & Pace& 83408 &  255\\ \hline
			1 & Argue contact &  47880 & 177\\ \hline     
			1 & Pace contact& 49110& 139\\ \hline
			\hline
			
			2 & Control & 19986 & 42\\ \hline
			2 & Argue& 70877 & 179\\ \hline
			2 & Pace& 68151 & 201\\ \hline
			2 & Argue contact & 44114 & 171\\ \hline     
			2 & Pace contact& 47086 & 75\\ \hline
			\hline
			
			3 & Control & 25863 &  88\\ \hline
			3 & Argue& 100079 &  425\\ \hline
			3 & Pace& 90942 & 506\\ \hline
			3 & Argue contact & 63382& 375\\ \hline     
			3 & Pace contact& 56851 & 234\\ \hline
			\hline
			4 & Control & 23205& 34\\ \hline
			4 & Argue& 60262 & 159\\ \hline
			4 & Pace& 47571 & 144\\ \hline
			4 & Argue contact & 44462& 146 \\ \hline     
			4 & Pace contact& 42059& 114\\ \hline
		\end{tabular}
		\caption{The number of tweets and tweets containing ``illegals'' in each phase and for each treatment group of the experiment.}
		\label{table:tweetcount}
	\end{table}

\section{Spillover Effect}
One source of contamination in our experiment could occur if a user retweeted the bot he followed, and then this retweet was seen by his follower who also followed a different bot.  This would cause the follower to receive treatments from two different bots, which is known as a spillover effect.  Though retweets happen very rarely in our experiment, we still wanted to make sure the spillover effect does not affect our results.

\begin{figure}
\centering
  \includegraphics[width=.9\linewidth]{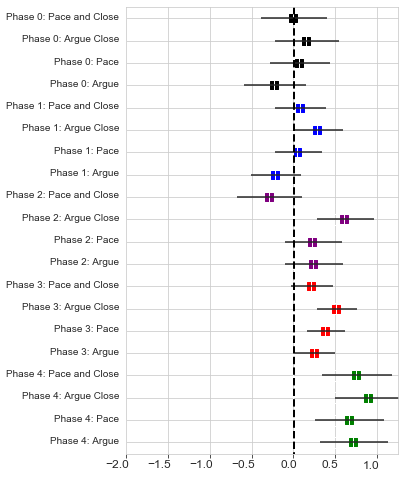}
\caption{Coefficient plots of regression with only refined users who did not experience the spill over effect. }
\label{fig:coef_refined}
\end{figure}

In total, 18 users retweeted the bots during the experiment. This results in 213 users (including the 18 retweeters) in the experiment who may have experienced the spill over effect. We excluded these users to cross-validate our result. We run logistic regression on both the whole user set, as well as the refined user set, and compare the results in Figures \ref{fig:coef} and \ref{fig:coef_refined}. As seen in the coefficient plots, the results are quite similar and we do not see any appreciable spillover effect in the regression coefficients.

\begin{figure}
\centering
  \includegraphics[width=.9\linewidth]{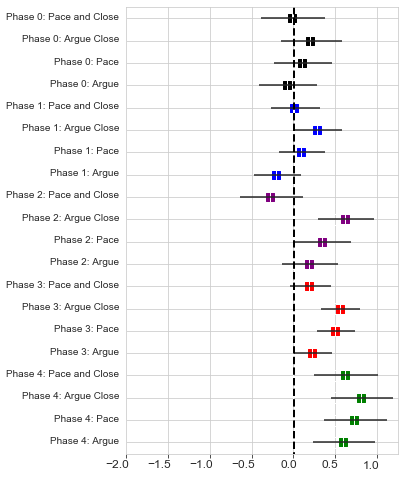}
\caption{Coefficient plots of regression with all users including those who may have experience the spillover effect.  These are the results in the main paper. }
\label{fig:coef}
\end{figure}

\end{document}